\documentclass[twocolumn,aps,superscriptaddress]{revtex4}

\usepackage{epsfig}
\usepackage{bm}
\usepackage{amssymb}
\usepackage[centerlast]{subfigure}

\newcommand{\dd}{{\mathrm{d}}}  
\newcommand{\HP}{{\mathit{HP}}}

\begin{document}
  
\title{Intermittent distribution of tracers advected by a compressible
random flow}

\author{J\'er\'emie\ Bec} \affiliation{Institute for Advanced Study, Einstein
  Drive, Princeton, New Jersey 08540, USA.}  \affiliation{Lab.\ G.-D.\ 
  Cassini, Observatoire de la C\^{o}te d'Azur, BP4229, 06304 Nice
  Cedex 4, France.}

\author{Krzysztof\ Gaw\c{e}dzki} \affiliation{CNRS, Laboratoire de Physique,
  ENS-Lyon, 46 All\'{e}e d'Italie, 69364 Lyon Cedex 7, France.}
\affiliation{Institute for Advanced Study, Einstein Drive, Princeton,
  New Jersey 08540, USA.}

\author{P\'eter\ Horvai} \affiliation{Centre de Physique Th\'{e}orique,
  \'{E}cole Polytechnique, 91128 Palaiseau Cedex, France.}
\affiliation{Institute for Advanced Study, Einstein Drive, Princeton,
  New Jersey 08540, USA.}  \affiliation{CNRS, Laboratoire de Physique,
  ENS-Lyon, 46 All\'{e}e d'Italie, 69364 Lyon Cedex 7, France.}

\begin{abstract}
  \noindent Multifractal properties of a tracer density passively
  advected by a compressible random velocity field are characterized. 
  A relationship is established between the statistical properties of
  mass on the dynamical fractal attractor towards which the
  trajectories converge and large deviations of the stretching rates
  of the flow. In the framework of the compressible
  Kraichnan model, this result is illustrated by analytical calculations 
  and confirmed by numerical simulations.
\end{abstract}

\maketitle

\noindent We are interested in the passive transport of a scalar
density by smooth-in-space compressible random flows in a
$d$-dimensional bounded domain $\Lambda$.  During transport, the
density develops strong inhomogeneities.  Our goal is to describe {
quantitatively} their fine structure arising at asymptotically large
times.  Compressible flows are physically relevant not only at large
Mach numbers. For instance, the dynamics of a suspension of
finite-size (inertial) particles in an incompressible flow may be
approximated in the limit of short Stokes times by that of simple
tracers in an effective compressible flow~\cite{ekr96,bff01}.  Another
application is to the advection of particles floating on the surface
of an incompressible fluid~\cite{SchumEckh}.

A smooth passive density field $\rho(t,\bm x)$ evolves in the velocity
field $\bm v(t,\bm x)$ according to the continuity equation
\begin{equation}
\label{eq:density}
  \partial_t \rho + \nabla\cdot(\rho\bm v)
=
  0
\end{equation}
which preserves the total mass that we shall assume for convenience
normalized to one.  If the initial density at time $t_0$ is uniform 
and if the flow is compressible and sufficiently
mixing, then the solution $\rho(t)$ of (\ref{eq:density}) will
approach, when $t_0$ tends to $-\infty$, a singular limit $\rho_*(t)$
which is a measure with support on the dynamical attractor towards
which the Lagrangian trajectories converge.  For random velocities, the
measure $\rho_*(t)$ and the dynamical attractor depend on the velocity
realization.  We shall consider stationary velocity ensembles where
the statistics of $\rho_*(t)$ does not depend on $t$. For convenience,
we shall restrict our study to $\rho_* \equiv \rho_*(0)$.  One expects 
the measure $\rho_*$ to have roughly a local product structure
with a smooth density along the unstable manifolds of the flow and 
a fractal-like structure in the transverse directions (such measures 
are called SRB~\cite{LSY}).

The \textbf{Lagrangian average} defined by
$$
  \langle F \rangle
\equiv
  \overline{
    \int_\Lambda
      F(\bm x \,|\, \bm v) \, \rho_*(\bm x \,|\, \bm v)
    \,\dd \bm x }
\, ,
$$
where the overline denotes the expectation with respect to the
velocity ensemble, samples points in $\Lambda$ according to 
the density $\rho_*$ of the asymptotic tracer distribution 
on the random attractor.  It determines
an invariant measure $\mu_*$ of the random dynamical system defined
on the product of the physical space $\Lambda$ and the space 
of the velocity realizations.

We are interested in the small-scale statistics of the mass
distribution associated to the measure $\rho_*$.  Denote by ${\cal
B}_r(\bm x)$ the ball of radius $r$ around the point $\bm x$ and
introduce the quantities
$$
  m_r(\bm x)
\equiv
  \int_{{\cal B}_r(\bm x)} \!\!\!\!  
    \rho_*(\bm y)
  \,\dd\bm y
\, , \quad 
  h_r(\bm x)
\equiv
  {\ln\, m_r(\bm x)\over \ln r}
\,.
$$
The \textbf{local dimension} at ${\bm x}$, defined if the limit exists
as $h(\bm x) \equiv \lim_{r\to0}\,h_r(\bm x)$, characterizes the
small-scale distribution of mass associated to the limiting density
$\rho_*$.  A more global assessment is provided by the
Hentschel-Procaccia spectrum for dimensions~\cite{HP,p97} of
$\rho_*\,$:
\begin{equation}
\label{HP}
  \HP(n \,|\, \rho_*)
\equiv
  \lim\limits_{r\to 0} \;
    \frac{\ln \int_\Lambda m_r(\bm x)^{n-1}\,\rho_*(\bm x)\,\dd\bm x}
         {(n-1) \ln r}
\end{equation}
for real $n$.  In particular, $\HP(0 \,|\, \rho_*)$ is the fractal
dimension of the support of $\rho_*$, $\HP(1 \,|\, \rho_*)$ is known
as the information (or capacity) dimension and $\HP(2 \,|\, \rho_*)$
as the correlation dimension.  For random flows, one may define the
\textbf{quenched} version of the Hentschel-Procaccia spectrum
$\HP_{\!\rm qu}(n) \equiv \overline{\HP(n \,|\, \rho_*)}$ and the
\textbf{annealed} version $\HP_{\!\rm an}(n)$ where the velocity
average is taken under the logarithm in the numerator on the right-hand 
side of (\ref{HP}). One has $\,(n-1)\HP_{\!\rm
qu}(n)\geq(n-1) \HP_{\!\rm an}(n)\,$ and $\,\HP_{\!\rm
qu}(1)=\HP_{\!\rm an}(1)\,$ (if exists). If the moments of $m_r(\bm x
\,|\, \bm v)$ with respect to the invariant measure $\mu_*$ exhibit the
small $r$ scaling
\begin{equation}
\label{scb}
  \langle\, m_r^n \,\rangle
\,\sim\,
  r^{\xi_n}
,
\end{equation}
then $\HP_{\!\rm an}(n+1)=\xi_n/n$.  There are
various aspects of behavior (\ref{scb}). First, a non-linear
dependence of the scaling exponents $\xi_n$ implies
\textbf{intermittency} in the mass distribution and, in particular,
the presence of non-Gaussian tails in the probability density function
(PDF) of $m_r$. Second, this behavior suggests that the PDFs of $h_r$
take, for small radii, the large deviations form $\,{\rm e}^{\,(\ln
r)\,S(h)}\,$ with the rate function $S(h)$ and the scaling exponents
$\xi_n$ related by the Legendre transform: $\,S(h) = \mathop{\rm max}_n
\,[\xi_n-nh]$. Third, the (annealed) fractal dimensions $f(h)$ of the
random sets on which the local dimension $h(\bm x)$ is equal to $h$
(the \textbf{multifractal spectrum}), is expected to be equal to
$h-S(h)$.

The aim of this letter is to relate quantitatively 
the dimensional
spectrum $\HP_{\!\rm an}(n)$ to the large deviations of the
\textbf{stretching rates} of the flow (sometimes also called 
finite-time Lyapunov
exponents). We note here that in \cite{bff01} a similar relation was 
discussed for $n=1$ and, in a qualitative way, for large $n$. 

Let $\bm X(t,\bm x)$ denote the Lagrangian trajectory passing at time
zero through point $\bm x$.  The matrix $\,W(t,\bm x) \equiv
\left({\partial_{j}X^{i}(t,\bm x)}\right)\,$ describes the flow
linearized around the trajectory $\bm X$. The eigenvalues of the
positive matrix $W^TW$ may be written, arranged in non-increasing
order, as $\,{\rm e}^{2t\sigma_1}, \dots, {\rm e}^{2t\sigma_d}\,$ with
$\sigma_i\equiv\sigma_i(\tau,\bm x|\bm v)$ called the stretching rates
of the flow. The limits
$$
  \lambda_i
\equiv
  \lim_{t\to\infty}\, \sigma_i(t)
$$
exist almost surely with respect to the invariant measure 
$\mu_*$ and are constant if the latter is ergodic. 
They define the Lyapunov exponents of the flow.
This is the content of the Multiplicative Ergodic Theorem~\cite{LArnold}.  
As argued in~\cite{bf99}, if the exponents $\lambda_i$ 
are all different then the PDF
of the stretching rates takes the large deviation form
\begin{equation}
\label{eq:ftle_largdev}
  \ {\rm e}^{-t\,H(\sigma_1,\dots,\,\sigma_d)}
  \,\, \theta(\sigma_1-\sigma_2)\dots\,\theta(\sigma_{d-1}-\sigma_d)
\end{equation}
for large $t$ with convex rate function $H$ attaining its minimum,
equal to zero, at $\sigma_i = \lambda_i$.

If the top Lyapunov exponent $\lambda_1$ is negative, then the
attractor measure $\rho_*$ is expected to be atomic with trivial mass
statistics.
For $\lambda_1>0$, Ledrappier and Young~\cite{ly88}
showed under rather general assumptions that the local dimension $h$
is almost surely equal to the Lyapunov dimension $d_L = j+\delta$
where $0 < \delta \leq 1$ is such that $\,\lambda_1 + \dots +
\lambda_j + \delta\lambda_{j+1} = 0$.  The above statement implies
that
\begin{equation}
\label{eq:dxi/dn|0}
  \HP_{\!\rm qu}(1)
=
  \HP_{\!\rm an}(1)
=
  {\dd \xi_n / \dd n} |_{n=0}
=
  d_L
\,.
\end{equation}
The Lyapunov dimension $d_L$ was introduced by Kaplan and
Yorke~\cite{ky79} and may be heuristically interpreted as the
dimension of objects keeping a constant volume during the time
evolution.  Eq.\,(\ref{eq:dxi/dn|0}) gives only partial information
about the mass scaling as compared to the full set of exponents
$\xi_n$.
\begin{figure}[b]
  \centerline{\includegraphics[width=0.3\textwidth]{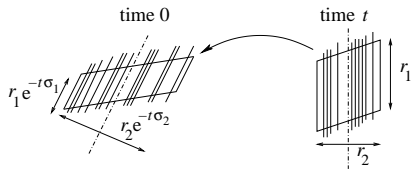}}
  \caption{\label{fig:mass_attract}
    Two-dimensional sketch of the backward-in-time evolution between
    times $t$ and $0$ of a small parallelogram of length $r_1$ in
    the direction of the unstable manifold and $r_2$ in 
    the direction perpendicular to it.}
\end{figure}

To relate the mass distribution in small balls to the fluctuations of
the stretching rates, we focus on the two-dimensional case.  The
expected smoothness of $\rho_*$ in the unstable direction leads us to
consider small parallelograms with one side of size $r_1$ parallel to
the unstable manifold and with extension $r_2$ in the direction
perpendicular to it.  We expect that the moments of the mass
$m_{r_1,r_2}$ in such parallelograms obey
\begin{equation}
\label{eq:rectscaling}
  \langle\,m_{r_1,r_2}^{\,n}\rangle
\sim
   r_1^{n} \, r_2^{\xi_n-n}
\end{equation}
for small $r_1,r_2$ as long as $\xi_n-n\geq 0$.  
When $\xi_n < n$, we
expect (\ref{eq:rectscaling}) to be replaced for 
$r_2 > r_1$ by
\begin{equation}
\label{eq:rectscaling-sat}
  \langle\, m_{r_1,r_2}^{\,n}\rangle
\sim
   r_1^{\xi_n} \, r_2^0
.
\end{equation}
This may be viewed as analogous to the stretching in one direction of a
fractal set of dimension $D$, while contracting it in the other one. 
In the expanding direction, the set will behave (down to a
scale depending on the stretching) as if projected on a line, so it
will have dimension $\min(1,D)$.  In the other direction it will have
the complementary dimension $D - \min(1,D) = \max(D-1,0)$.

Let us consider such a parallelogram at $t>0$ and let us look at 
its pre-image at time zero. While the direction parallel to the unstable 
manifold is exponentially contracted backward-in-time with a rate given 
by the largest stretching rate $\sigma_1$, the other direction typically
expands with an exponential rate $\sigma_2$.  Hence the time-zero pre-image
of the original parallelogram is (approximately) another parallelogram 
as sketched in
Fig.~\ref{fig:mass_attract}. Conservation of mass and stationarity
of the statistics lead to the relation 
$$
  \langle\, m_{r_1,r_2}^{\,n}\rangle
\approx
  \langle\, m_{r_1{\rm e}^{-t\sigma_1}\,,\,r_2{\rm e}^{-t\sigma_2}}^{\,n}
\rangle.
$$
If there is a sufficiently rapid loss of memory in the Lagrangian
dynamics (i.e.\ if the invariant measure is sufficiently mixing)
then the expectation on the right-hand side should factorize for large $t$
(such a factorization holds for all $t$ in the Kraichnan model discussed 
below). In such situation, using the large deviation form 
(\ref{eq:ftle_largdev}) of the PDF of the stretching rates, we infer that 
$$
  \langle\, m_{r_1,r_2}^{\,n} \rangle
\sim
  \!\int_{\sigma_1\geq\sigma_2}\!\!\!\!\!\!
    \langle\,
      m_{r_1{\rm e}^{-t\sigma_1}\,,
      \,{r_2\rm e}^{-t \sigma_2}}^{\,n}
    \rangle\,
  {\rm e}^{-t H(\sigma_1,\dots,\,\sigma_2)}\;\dd\sigma_1
\dd\sigma_2\,.
$$
Consistency of the above relation with the scaling
(\ref{eq:rectscaling}) requires that
$$
  1
\sim
  \int_{\sigma_1\geq\sigma_2} \!\!\!\!\!\!\!
  {\rm e}^{-t\,[n\sigma_1+(\xi_n-n)\sigma_2 +
                H(\sigma_1,\,\sigma_2)]}\,\,\dd\sigma_1\dd\sigma_2\,.
$$
Since $t$ is assumed large, a saddle-point argument implies
then the following relation between the scaling exponents $\xi_n$ and 
the rate function $H$ of the stretching rates:
\begin{equation}
\label{eq:form2d}
  \min_{\sigma_1\geq\sigma_2} \left[
    n\sigma_1 + (\xi_n-n)\sigma_2 +
    H(\sigma_1,\,\sigma_2)
  \right]
=
  0
.
\end{equation}
Alternative formulations are
$$
  \xi_n
=
  n -
  \!\!\!
  \max\limits_{
          \begin{array}{c}
            \raisebox{3pt}[0pt][0pt]
                     {$\scriptstyle \sigma_1 \geq \sigma_2$}
          \\
            \raisebox{6pt}[0pt][0pt]{$\scriptstyle \sigma_2 < 0$}
          \end{array}
       }\!
    \frac{1}{\sigma_2}
    \left[
      n\sigma_1 + H(\sigma_1,\,\sigma_2)
    \right]
=
  \min\limits_{h\geq0} \, [h n + S(h)]
$$
for $\,S(h)=\min_{\sigma>0}\,{\sigma}^{-1} H((h-1)\sigma,-\sigma)$.
These formulae are valid for $\xi_n \geq n$.  Similarly, from
(\ref{eq:rectscaling-sat}) we may get the formula valid for $\xi_n
\leq n$
\begin{equation}
\label{eq:form2d-sat}
  \xi_n
=
  -\max_{
           \begin{array}{c}
             \raisebox{3pt}[0pt][0pt]
                      {$\scriptstyle \sigma_1 \geq \sigma_2$}
           \\
             \raisebox{6pt}[0pt][0pt]{$\scriptstyle \sigma_1 < 0$}
           \end{array}
        }
    \frac{1}{\sigma_1}
    H(\sigma_1,\,\sigma_2)
.
\end{equation}

It is easily checked from (\ref{eq:form2d}) that $(\dd\xi_n/\dd
n)|_{n=0} = 1 - \lambda_1/\lambda_2$ which coincides, in our settings,
with formula (\ref{eq:dxi/dn|0}).  Also from (\ref{eq:form2d}) one may
deduce that
$$
  \lim_{t\to\infty}
    \frac{1}{t}
    \left\langle
      \left| \bm R(t) \right|^{-\xi_1}
    \right\rangle
=
0
,
$$
for $\bm R(t)=W(t,\bm x)\bm R_0$, meaning that the generalized
Lyapunov exponent of order $-\xi_1$ vanishes. This is a known result
for stochastic flows of diffeomorphisms, see~\cite{b91}.
Relation (\ref{eq:form2d}) can be extended to dimensions 
higher than two using similar arguments.

To illustrate the small-scale properties of the mass distribution, we
focus in the sequel on transport by a random compressible velocity
field chosen in the framework of the Kraichnan model~\cite{Kr68}.  For
the domain $\Lambda$ we take a periodic box.  The velocity $\bm v$ is
taken centered Gaussian, with covariance
$$
  \overline{v^i(\bm x+\bm\ell,t+\tau)\, v^j(\bm x,t)} =
  2(D_0 \delta^{ij} - d^{ij}(\bm\ell))\,\,\delta(\tau),
$$
where, for small separations $\bm\ell$, the function $d^{ij}$ satisfies
\[
  d^{ij}(\bm \ell)
=
    \frac{\mathrm{D}_1}{2}\left[(d+1-2\wp)\,\delta^{ij}\ell^2
  + 2(\wp d-1) \,\ell^i\ell^j
  + o(\ell^2)\right]
\]
assuring local isotropy.  The parameter $\wp$, called the
compressibility degree, is chosen in the interval~$[0,1]$. Its extreme
values correspond, respectively, to incompressible and to potential
velocity fields. For such velocity ensembles, the distribution of the
stretching rates is explicitly known~\cite{bf99,fgv01}. The
corresponding rate function takes the simple form
$$
  H = \frac {1}{\mathrm{C}_1} \Big[ \sum _{i=1}^{d}
  \,(\sigma_i-\lambda_i)^{2} + \mathrm{C}_2 \Big( \sum _{i=1}^{d}
  \,(\sigma_i-\lambda_i)\Big)^{2} \Big].
$$
$\mathrm{C}_1 \equiv 4 \mathrm{D}_1\,(d + \wp\,(d - 2))$,
$\mathrm{C}_2 \equiv (1 - \wp\,d)/(\wp\,(d - 1)\,(d + 2))$.  The
Lyapunov exponents are $\lambda_j = \mathrm{D}_1
[d\,(d-2j+1)-2\wp(d+(d-2)j)]$.  In the two-dimensional case, the
largest Lyapunov exponent is negative when $\wp>1/2$.  For $\wp<1/2$,
the exponents for the mass distribution obtained from
(\ref{eq:form2d}) and (\ref{eq:form2d-sat}) read
\begin{equation}
\label{eq:expokraich}
  \xi_n
=
  \left\{\begin{array}{cl}
    \frac{2n + \sqrt{(1+2\wp)^2 - 8\wp n}}{1+2\wp} -1 &
    \ \mbox{ if }\quad n \le n_{\rm cr}\,,
\\[10pt]
    \xi_{\infty} &
    \ \mbox{ if }\quad n \ge n_{\rm cr}
  \end{array}\right.
\end{equation}
where the critical moment $n_{\rm cr}$ and the saturation exponent
$\xi_{\infty}$ are given by
\begin{eqnarray}
  \left\{\begin{array}{rcl} n_{\rm cr} &=&
  \frac{1}{2}\sqrt{{1+\frac{1}{2\wp}}} \\
  \xi_{\infty} &=& 2 n_{\rm cr} -1
\end{array} \right.  &\quad\mbox{if}\quad
  0<\wp\le{1/6}\,,
\nonumber
\\[5pt] n_{\rm cr} =
  \xi_{\infty} = \xi_1 =\frac{2-4\wp}{1+2\wp} &\quad\mbox{if}\quad
  {1/6}\le\wp<\frac{1}{2}\,. 
\label{eq:defsat2}
\nonumber
\end{eqnarray}
The two different behaviors are illustrated in Fig.~2.  In both cases,
the events contributing to the saturation of the exponents are those
for which a mass of order unity is concentrated inside the small ball.

\begin{figure}[htbp]
  \centerline{\includegraphics[width=0.2\textwidth]{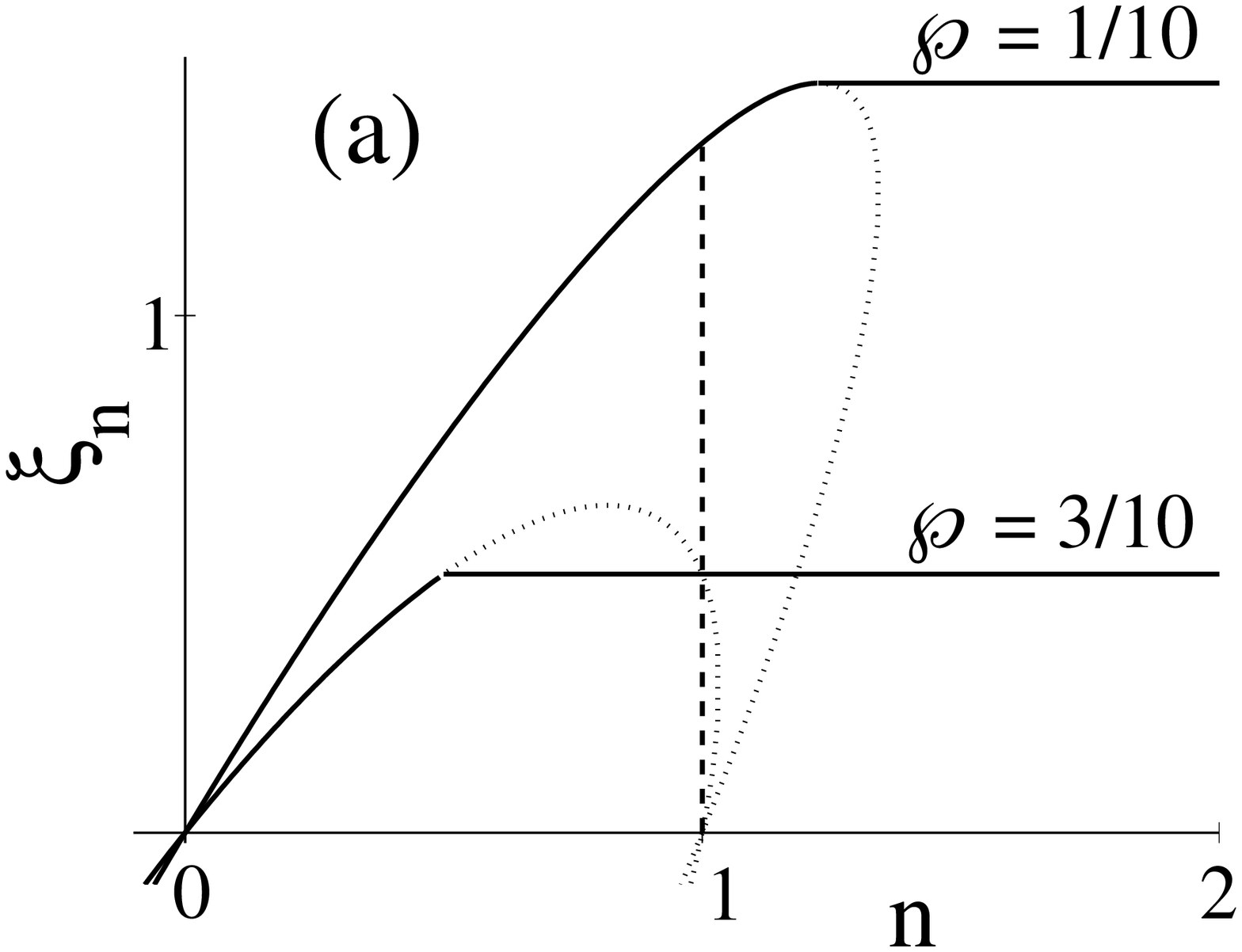} \hfill
    \includegraphics[width=0.22\textwidth]{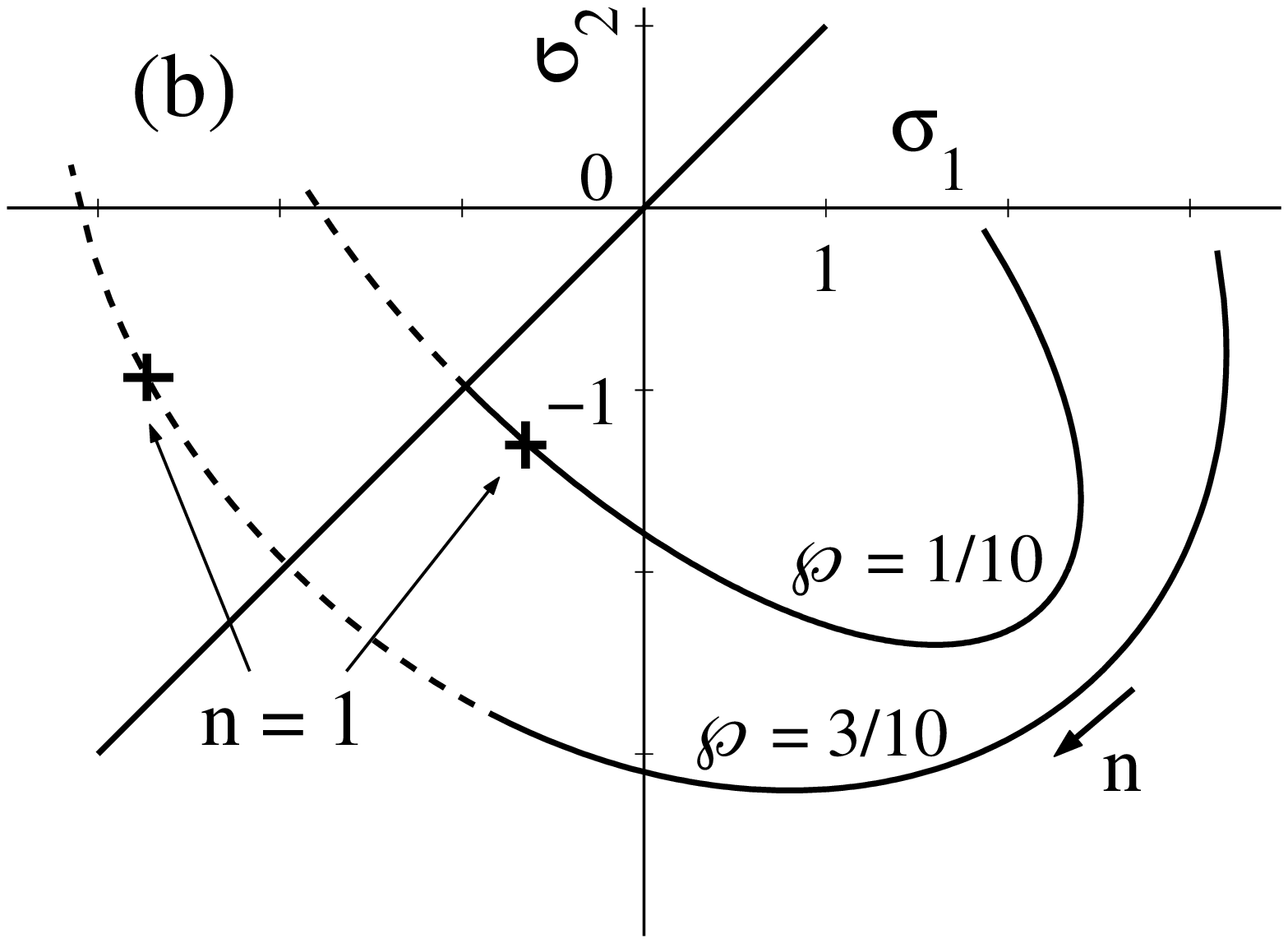}}
  \caption{\label{fig:saturation} (a)~Ellipses in the
    $(n,\,\xi_n)$ plane corresponding to the unconstrained minimum in
    (\ref{eq:form2d}).  Dotted parts refer to situations when either
    the minimum is reached for $\sigma_1 < \sigma_2$ or $\xi_n < n$.
    Two values of $\wp$ are represented to illustrate both the case
    $\wp < 1/6$ and $\wp > 1/6$.  (b)~Location where the minimum is
    reached in the $(\sigma_1,\sigma_2)$ plane.}
  \vspace{-2ex}
\end{figure}

\begin{figure}[htbp]
  \centerline{
    \subfigure[\label{fig:exponents_P_0.1} $\wp = 1/10$]{%
      \includegraphics[width=0.22\textwidth]{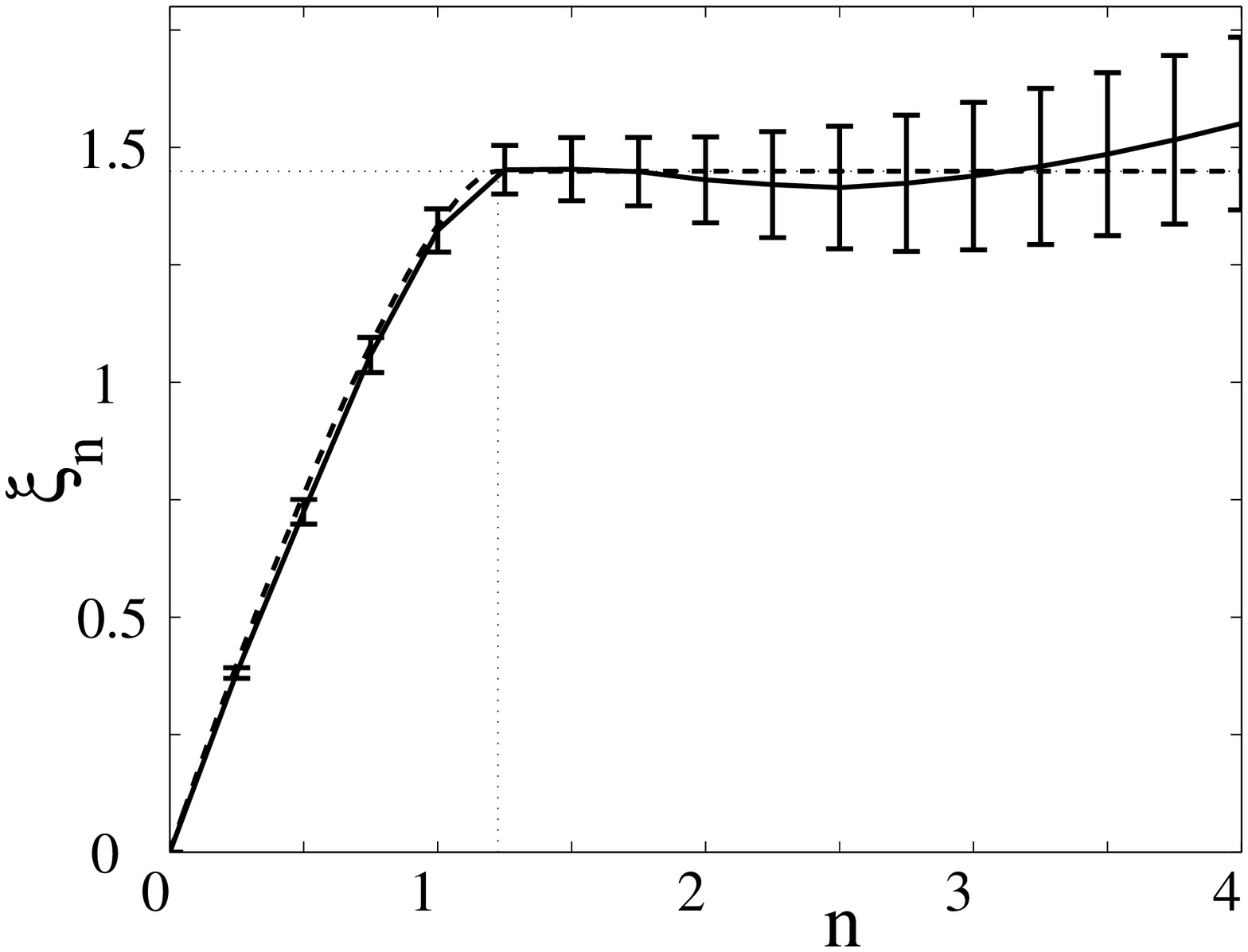}}\hfill
    \subfigure[\label{fig:exponents_P_0.3} $\wp = 3/10$]{%
      \includegraphics[width=0.22\textwidth]{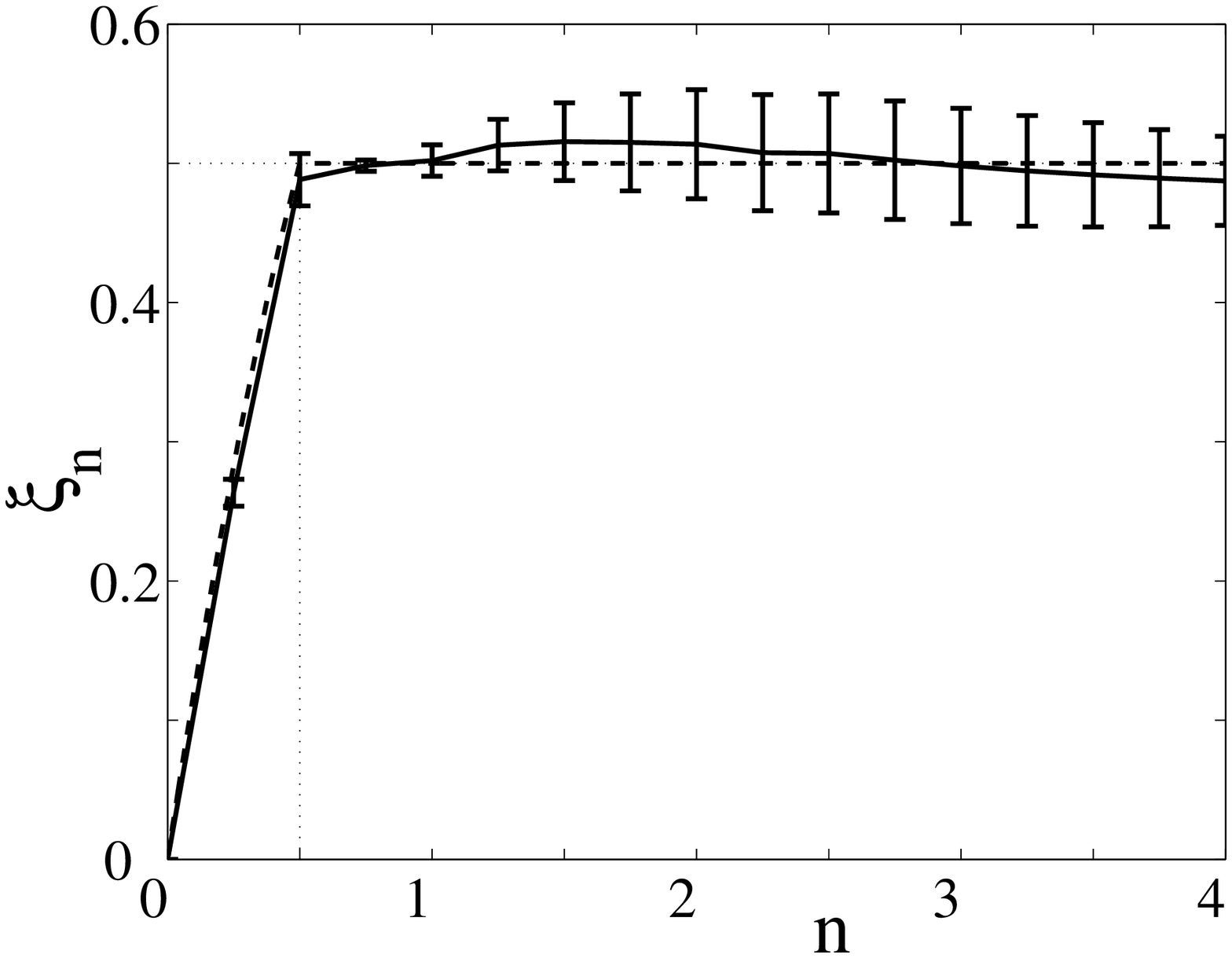}}
  }
  \caption{\label{fig:scaling1} Scaling exponents $\xi_n$ for the mass
        distribution associated to the advection of a density field by
        a Kraichnan velocity field.  For two different degrees of
        compressibility, the exponents obtained match those predicted
        by theory which are represented as dashed lines.}
\end{figure}

Numerical simulations confirm the values of the scaling exponents
obtained explicitly for the compressible Kraichnan model.  To
distinguish the two cases, two different values of the compressibility
degree are investigated ($\wp = 1/10<1/6$ and $\wp = 3/10>1/6$).  The
velocity field $\bm v$ is generated by the superposition of nine
independent Gaussian modes and the density is approximated by
considering a large number of Lagrangian tracers.  The exponents
obtained numerically after averaging over $10^5$ turnover times and
for $N=10^5$ tracers are shown in Fig.~\ref{fig:scaling1}.  Although
statistical convergence of the average is quite slow, these expensive
simulations are in a rather good agreement with the theory, in
particular with the saturation of the exponents after the critical
order.

\begin{figure}[htbp]
  \centerline{\includegraphics[width=0.24\textwidth]{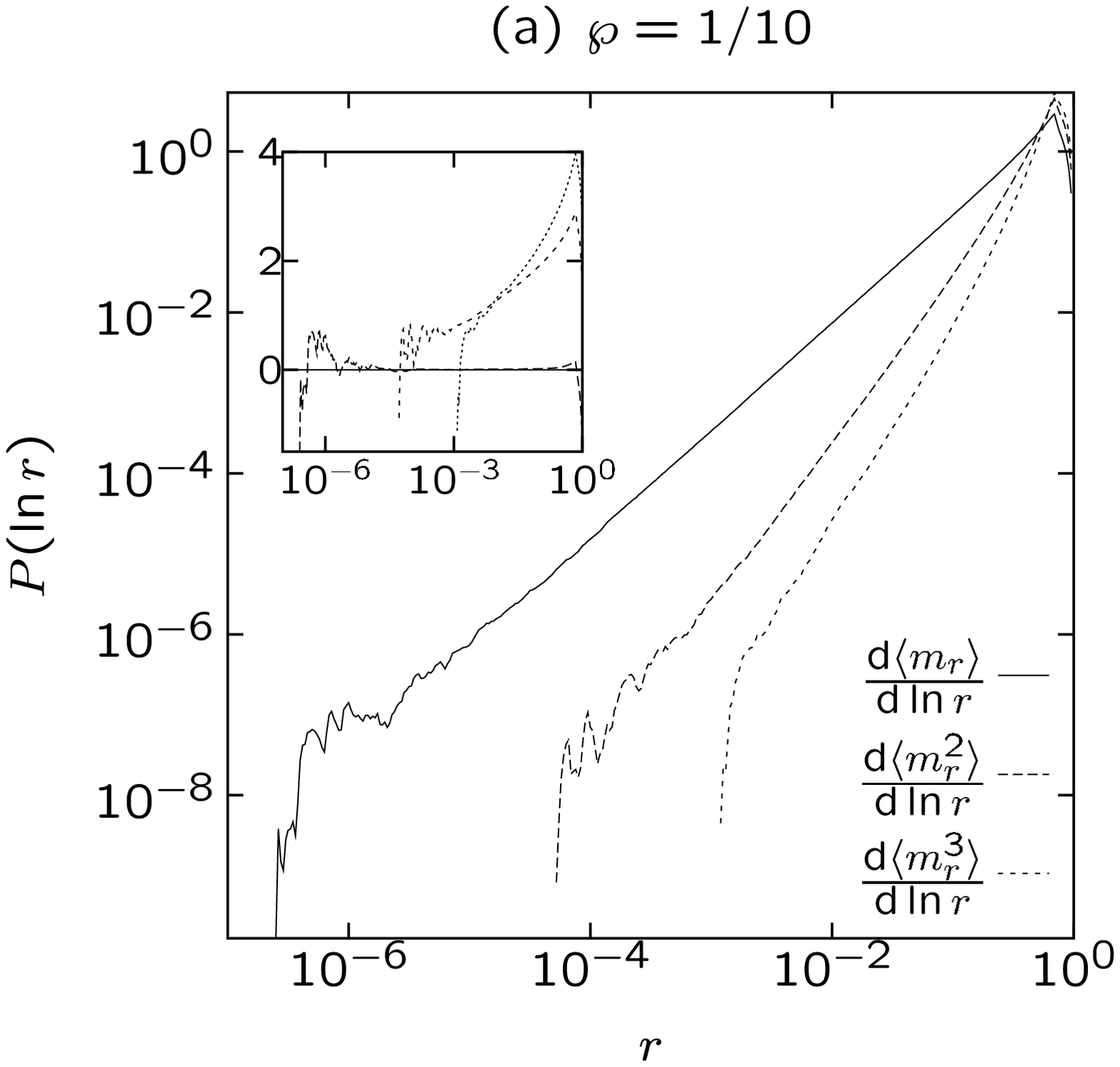}
              \includegraphics[width=0.24\textwidth]{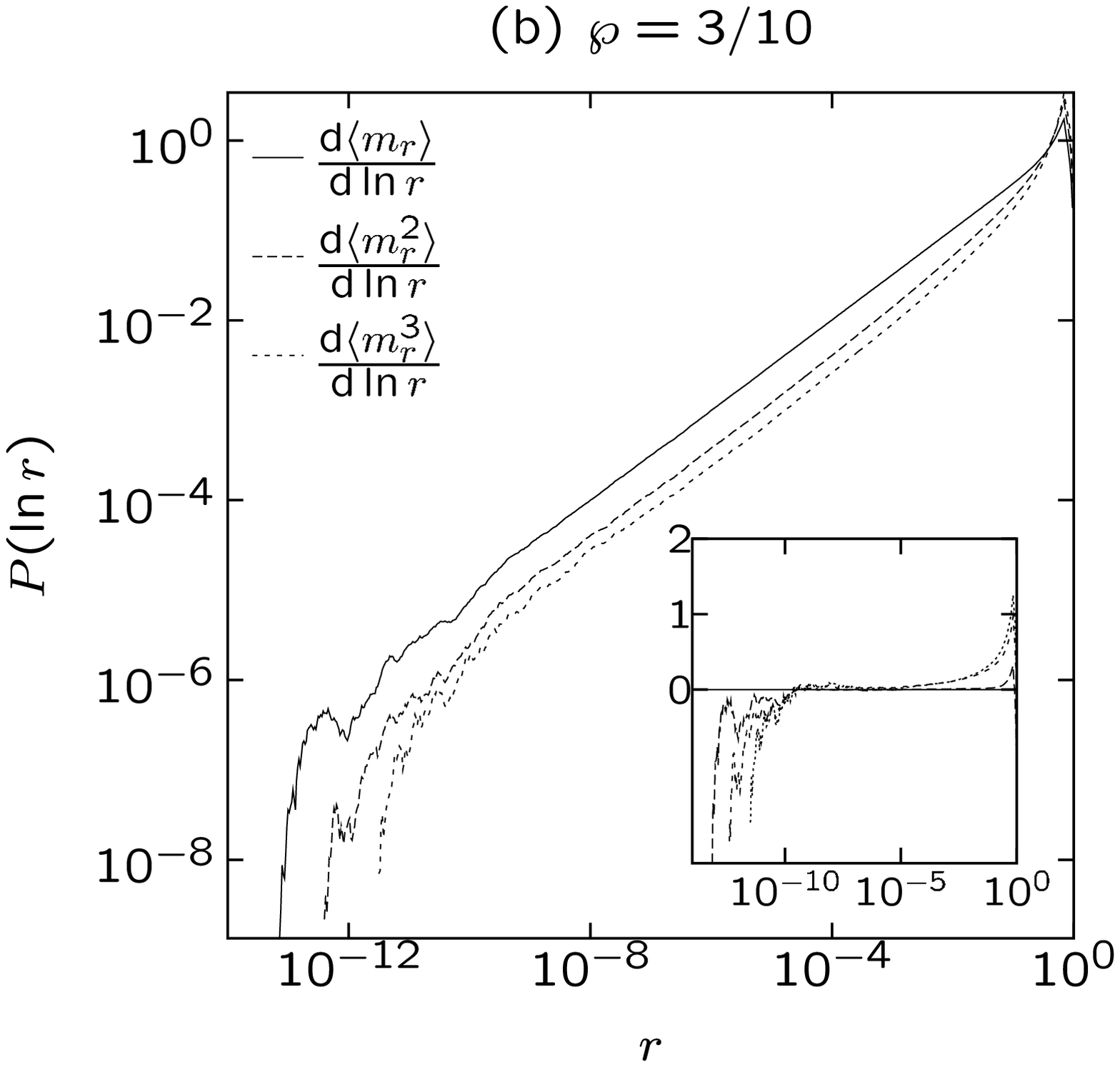}
              \hspace{1.5em}}
\vskip -3cm
  \caption{\label{fig:scaling2}  For $n+1$ particles ($n=1,\,2$ and
    $3$) the distribution of the maximum of the distances of the
    reference particle to the other $n$ particles is represented.  It
    can be shown that this PDF is proportional to $\,\dd \langle
    m_r^n \rangle/\dd \ln r$. \,The insets show the 
difference with small-$r$ asymptotes with slopes
given by the theoretical exponents. 
}
\end{figure}

The determination of the exponents may be improved for positive
integer orders $n$ by considering only $n+1$ particles.  This method
allows to perform very long time averages (here of the order of $10^8$
turnover times) required for good convergence of the statistics at
small scales.  As shown in Fig.~\ref{fig:scaling2}, for $\wp = 3/10$,
the moments clearly scale over several decades.  For the lower value
of compressibility $\wp = 1/10$, convergence is slower because of the
smaller probability for the tracked particles to come close together.

Finally, let us mention that, for integer orders, the exponent
$\xi_n$ for the Kraichnan model can be linked to the homogeneity
degree of the stationary single-time density correlation function
$${\cal F}_{n+1} \equiv \overline{\rho({\bm 0})\rho({\bm
x}_1)\cdots\rho({\bm x}_n)}\,.$$  The $n^{\rm th}$ order
moment of $m_r$ may be written as
\begin{equation}
\label{eq:masstodensity}
  \left\langle\, m_r^n \right\rangle\hspace{-0.1cm}
\ =
  \displaystyle \int_{{\cal B}_r({\bm 0})\times
    \cdots\times{\cal B}_r({\bm 0})} \!\!\!\!\!\!\!\!\!\!\!\!\!\!\!
  \!\!\!\!\!\!\!\!\!\!\!\!\!\!\!\!\!\! 
  {\cal F}_{n+1} ({\bm 0},{\bm x}_1,
  \dots,{\bm x}_n)\,\,\dd{\bm x}_1\dots \dd{\bm x}_n\,.\ \quad
\end{equation}
In the stationary regime of the Kraichnan model ${\cal F}_{n+1}$ is a
zero mode of the operator
$$
  M_{n+1}^\dag
\equiv \!\!\!\!
  \sum_{0\le k,\ell\le n} 
    \partial_{x_k^i} \partial_{x_\ell^j}
    \left[ \left(   
      D_0\delta^{ij} - d^{ij}(\bm x_k -\bm x_\ell )
    \right)\,\cdot\,\,\right]
$$
We need not write these zero modes explicitly.  The homogeneity degree
of the isotropic solution of lowest degree can be found simply by
requiring its positivity and imposing on it certain continuity and
integrability conditions.  The branch $n < n_{\rm cr}$ in
(\ref{eq:expokraich}) is obtained by requiring the solution to be
continuous at ${\bm x}_1 = \ldots = {\bm x}_n$.
On this subspace, $M_{n+1}^\dag$ is degenerate, owing to the fact that
collinear points remain collinear when transported by the (linearized)
flow.  If in addition we restrict $M_{n+1}^\dag$ to the rotationally
invariant sector, it becomes an ordinary second-order homogenous
differential operator.  It has two scaling solutions. The one with
exponent $\xi_n - nd$ gives, through (\ref{eq:masstodensity}), the
non-saturated ($n < n_{\text{cr}}$) branch of
(\ref{eq:expokraich}). Such a solution breaks down when the
corresponding zero mode ceases to be integrable at small ${\bm x}_i$
which may occur if the restriction of the zero mode to the collinear
sector is not integrable near 0, i.e. $\xi_n - nd \leq -1$.  This
gives $n_{\rm cr}$ and $\xi_\infty$ in the case $\wp < 1/6$.  Another
scenario is if the zero mode has a non-integrable singularity around
the collinear geometry.  This gives $n_{\rm cr}$ and $\xi_\infty$ in
the case $\wp > 1/6$.

An important open question not touched upon by this paper concerns the
case of rough-in-space velocity fields appearing in the limit of very
high Reynolds numbers.  That problem cannot be formulated in terms of
stretching rates but the relationship with density correlations and
zero modes still holds.

We are grateful to D.~Dolgopyat, U.~Frisch, K.~Khanin, Y.~Le~Jan and
O.~Raimond for interesting and motivating discussions. 
J.B.\ acknowledges the support of the National Science Foundation 
under agreement No.\ DMS-9729992 and of the European Union under contract
HPRN-CT-2000-00162.  K.G.\ thanks the von Neumann Fund at IAS in Princeton 
for the grant. Part of the numerical simulations were performed in 
the framework of the SIVAM project at the Observatoire de la
C\^{o}te d'Azur.

When this work was essentially finished, we learned from A.~Fouxon
that he has also derived the relationship (\ref{eq:form2d}) between
the scaling exponents of mass and the large deviations of the
stretching rates.

\end{document}